\documentclass[twocolumn,preprintnumbers,amsmath,amssymb,showpacs]{revtex4}
\usepackage{graphicx,epsfig}   
\usepackage{amsmath}
\usepackage[latin1]{inputenc}
\usepackage{psfrag}
\usepackage{plain}
\usepackage{tabularx}

\renewcommand{\vec}[1]{\boldsymbol{\rm #1}}

\newcommand{\rhosand}{\rho_{\text{sand}}}
\newcommand{\rhoair}{\rho_{\text{air}}}

\newcommand{\veff}{v_{\text{eff}}}

\def\frac#1#2{{\mathinner{#1 \over #2}}}

\def\dmath#1#2{
$$\lineskiplimit=1000pt \advance\lineskip by #1\jot
\mathsurround=0pt \tabskip=0pt plus 1000pt
\everycr{\noalign{\penalty\interdisplaylinepenalty}}
\halign to \displaywidth{
\hfil$\displaystyle{##}$\tabskip=0pt&%
\hfil $\displaystyle{{}##{}}$\hfil &%
$\displaystyle{##}$\hfil \tabskip=0pt plus 1000pt minus 1000pt&%
\refstepcounter{equation}\label{##}\llap{(\theequation)}\tabskip=0pt\cr
\noalign{\ifdim \prevdepth>-1000pt \vskip -#1\jot\fi}
#2\crcr}$$}

\def\crl#1{\crcr\noalign{\unpenalty\penalty 10000
\nointerlineskip \vbox to 0pt {
\dimen0=\lineskip \vskip \dimen0 minus 1000pt \hbox to \displaywidth{%
\hfil \refstepcounter{equation}\label{#1}(\theequation)}
\vskip 0pt minus 1000pt} \penalty 10000}}

\def\mbf#1{\mathchoice{\hbox{\boldmath $\displaystyle #1$}}
        {\hbox{\boldmath $\textstyle #1$}}
        {\hbox{\boldmath $\scriptstyle #1$}}
        {\hbox{\boldmath $\scriptscriptstyle #1$}}}

\def\ind#1{{\hbox{\tiny #1}}}

\begin{document}

\title{Breeding and Solitary Wave Behavior of Dunes}

\author{O. Dur\'an $^1$, V. Schw\"ammle $^{1,2}$ and H. Herrmann $^{1,3}$}
\affiliation{$^1$ ICP, University of Stuttgart,
70569 Stuttgart,  Germany}
\affiliation{$^2$ Instituto de F\'isica, Universidade Federal Fluminense, Av.
Litor\^anea s/n, Boa Viagem; Niter\'oi 24210-340, RJ, Brazil.}
\affiliation{$^3$ Departamento de Física, Universidade Federal
do Ceará, 60455-970 Fortaleza, Brazil}

\date{\today}

\begin{abstract}

Beautiful dune patterns can be found in deserts and along
coasts due to the instability of a plain sheet of sand under the action of the
wind. Barchan dunes are highly mobile aeolian dunes found in areas of low sand
availability and unidirectional wind fields. Up to now modelization mainly
focussed on single dunes or dune patterns without regarding the mechanisms of
dune interactions. We study the case when a small dune bumps into a bigger one.
Recently Schwammle \& Herrmann ({\em Solitary wave behavior of sand dunes.} 
Nature {\bf 426}, 610 (2003)) and Katsuki {\em et al.}({\em  Collision dynamics
of two Barchans dunes simulated by a simple model.} cond-mat 0403312 (2004)) 
have shown that under certain
circumstances dunes can behave like solitary waves. This means that they can
``cross'' each other which has been questioned by many researchers
before. In other cases we observe coalescence, i.e. both dune merge into one,
breeding, i.e. the creation of three baby dunes at the center and horns of a Barchan, or
budding, i.e. the small dune, after ``crossing'' the big one, is unstable and
splits into two new dunes.

\end{abstract}


\maketitle

\section{Introduction}

\begin{figure*}[htb]
  \begin{center}
    \includegraphics[width= 1.0 \textwidth]{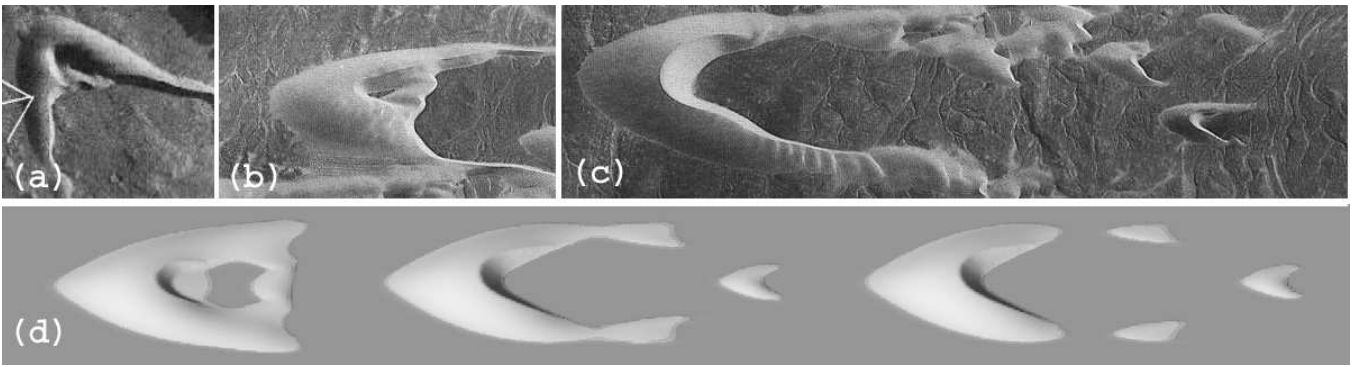}
\caption{Examples of the breeding process in two Barchan fields, Namibia (a) and Peru (b), (c),
and in numerical simulations (c).}
    \label{fig.1}
\end{center}
\end{figure*}

We observe many different dune patterns in nature, as for example
longitudinal, transverse, star and Barchan dunes. In regions where the wind
blows mainly from the same direction sand availability determines the dune
pattern. At high sand disposal transverse dunes dominate in the fields. They
seem to be translationally invariant so that the lateral sand
flux can be neglected. When less sand is available, Barchan dunes appear
(Fig.~\ref{fig.1}). These are highly mobile having the form of a crescent moon.
Their velocity can reach up to several tenths of meters per year and is
proportional to the reciprocal height,
meaning that smaller dunes are faster than large ones.
The surface of a Barchan
can be divided into different sections: the windward side, the slip face after
the brink and the horns from which sand can leave the dune. Barchan dunes of
different sizes are not perfectly shape invariant and there exists a minimal
height of 2-3 meters below which they are not stable.
These observations result from field measurements that have been made over the
last decades \cite{Bagnold,Finkel,Coursin,Hesp,Jimenez,Sauermann00,Sauermann03}.
Still many questions on dune dynamics remain open. Due to the large time scales
involved in dune formation, full evolution of a dune is difficult to assess
through measurements. Attempts have been made to get more insight through
numerical calculations. Recently, several numerical models have been proposed in
order to explain dune morphology and formation
\cite{Wippermann,Zeman,Fisher,Stam,Nishimori,vanBoxel,vanDijk,Momiji,
Sauermann01,Kroy,Andreotti,Hersen,Schwammle1,Schwammle2}. They have to deal with
the calculation of the turbulent wind field, the saltation sand flux over the
windward side and the avalanches going down the slip face. Up to now
modelization mainly focussed on single dunes or dune patterns without regarding
the mechanisms of dune interactions.

Recently, Besler found small Barchans at the downwind side of big ones and
concluded that Barchan dunes could behave like solitons
\cite{Besler97,Besler}.  This means that they would behave like solutions of
non-linear equations, for  example those describing waves in shallow water,
which propagate through each  other without changing their shape \cite{Lamb}.
As an example see  Fig.~\ref{fig.1}. In (a) and (b) a small Barchan is apparently
ejected from the main dune, whereas, in (c) small dunes emerge from the horns.
Note the similarity with the snapshots of a collision simulation depicted in (b).
Similar occurrences can be found in experiments with sub-aqueous Barchans \cite{endo}.
Nevertheless, most  researchers believe that if a small Barchan hits a bigger
one, it will be  completely absorbed. This is motivated by the fact that a sand
formation cannot  cross the slip face of a dune without being destroyed.
Therefore the description  of Barchan dunes as solitons has found very little
support up to now, until  Schw\"ammle et al. found that dunes can behave as
solitary waves under certain  conditions \cite{Schwammle}. They show that, due
to mass exchange, a big Barchan colliding with a smaller one placed behind, may
decrease its height until it becomes smaller, and therefore, faster, than the
previous one, and leaves. Meanwhile the initially smaller dune increases its
height becoming bigger and slower, in such a way that it looks as if the
smaller dune crosses the big one. This situation was referred as solitary wave
behavior. Katsuki et al. \cite{katsuki} also have obtained solitary wave
behavior for
coaxial and offset collisions of two sub-aqueous Barchans.

In the following we use a minimal model for dunes to develop the
morphological phase diagram for the coaxial collision between two Barchans. We 
also
characterize the transition between the different regimes using two order
parameters and show a possible size selection mechanism in Barchan fields.

\section{Model}

Our model \cite{Kroy,Sauermann01,Schwammle1} consists of three coupled equations of
motion calculating the shear stress of the wind field, the sand flux
, the avalanche flux and the resulting change
of the topography using mass conservation. The shear stress of the wind is
obtained from the perturbation of the air flow over a smooth hill using the well
known logarithmic velocity profile of the atmospheric boundary layer
\cite{Prandtl}. The shear velocity describes the strength of the wind. At the
windward side of a Barchan dune the sand grains are transported by the so called
saltation mechanism \cite{Bagnold}. Below the big vortex of the wind field
behind the brink the air cannot maintain sand transport. So the grains are
deposited behind the brink in the lee zone of the Barchan until the surface
reaches the angle of repose on which the sand flows down the slip face through
avalanches.

The simulations are carried out with a completely unidirectional and constant
wind source. In every iteration the horizontal shear stress $\vec{\tau}$ of the
wind, the saltation flux
$\vec q$ and the flux due to avalanches are calculated. The time scale of these
processes
is much shorter than the
time scale of changes in the dune surface so that they are treated to be
instantaneous.
We perform all simulations using open boundary conditions with a  constant influx.
In the following the different steps at every iteration are explained.

\paragraph{The air shear stress $\tau$ on the ground:}

The shear stress is computed according to an analytical work describing the
perturbation of the ground shear stress by a low hill or dune
\cite{Weng}.  This perturbation is given by:
\begin{eqnarray}
\label{eq:shear}
\tilde{\hat\tau}_x &=& \frac{\tilde h_{s}\;k_x^2}{|\mbf k|}
\frac2{U^2(l)}
\left\{ - 1 + \left(2 \ln\frac l{z_0} + \;\frac{|k|^2}{k_x^2}\right)
      \sigma \;\frac{K_1(2\sigma)}{K_0(2\sigma)} \right\},\nonumber \\
\tilde{\hat\tau}_y &=& \frac{\tilde h_{s}\;k_x\,k_y}{|\mbf k|}
\frac2{U^2(l)}
  \,2\sqrt{2}\,\sigma K_1\!\big(2\sqrt{2}\,\sigma\big)\,,
\end{eqnarray}
\noindent where $\sigma = \sqrt{i\,L\,k_x\,z_0/l}$.

Here, $K_0$ and $K_1$ are modified Bessel functions.  $k_x$ and $k_y$ are the
components of the wave vector $\mbf k$, the coordinates in Fourier space.
$\tilde{\hat\tau}_x$ and $\tilde{\hat\tau}_y$ are the Fourier-transformed
components of the shear stress perturbation in wind direction and in transverse
direction.  $\tilde h_{s}$ is the Fourier transform of the height profile,
$U$ is the vertical velocity profile which is suitably nondimensionalised, $l$
the depth of the inner layer of the flow, and $z_0$ the roughness length which
takes into account saltation.  $L$ is a typical length scale of the hill or
dune and is given by a quarter of the mean wavelength of the Fourier
representation of the height profile. In order to take into account the shear 
stress strong reduction in the lee side of the hill, where the linear models
overestimate it \cite{hewer}, we multiplied the imaginary part 
of $\tilde{\hat\tau}_x$ by the fenomenological constant $1.5$.

It has to be taken into account that
the flow separates at the brink of the slip face of a dune.  This is done by
assuming an idealized ``separation bubble'', a region inside which there is no
flow and outside of which the air flows as over a shape of the combined dune
and bubble.  Each slice in wind direction of the bubble is modelled by a
third-order polynomial so that in the case of a barchan the region between the
horns is inside the bubble.  The height profile $h_{s}$ in (\ref{eq:shear})
is the profile of the dune including the separation bubble.

\paragraph{The saltation flux $q$:}

From the shear stress, the modification of the air flow due to the
presence of saltating grains is accounted for.  This results in an effective
wind velocity driving the grains.
\begin{equation}
\label{eq:3d:v_eff_of_tau_g0}
v_{\text{eff}} =
 \frac{u_{*t}}{\kappa} \left(\ln{\frac{z_1}{z_0}} + 2
 \left(\sqrt{1 + \frac{z_1}{z_m}\left( \frac{u_{*}^2}{u_{*t}^2} - 1\right)} - 1 \right) \right),
\end{equation}
\noindent where $u_* = \sqrt{\tau / \rhoair}$

The shear stress $\mbf\tau$ results from (\ref{eq:shear}) through
$\mbf\tau=\mbf\tau_0 + |\mbf\tau_0|\,\mbf{\hat\tau}$, where $\mbf\tau_0$ is the
shear stress over a flat plane.  $z_0$ is the roughness length of
the sand excluding the effect of saltation, and $\kappa\approx 0.4$ is von
K\'arm\'an's constant.  $z_m$, the mean saltation height, and $z_1$ are
parameters of the model.

The next step is the computation of the typical velocity of the saltating
grains.  It is determined by the balance between the drag force acting on the
grains, the loss of momentum when they splash on the ground, and the downhill
force.  It is computed by solving the following quadratic vector equation
numerically:
\begin{equation}
  \label{eq:u}
   \frac{3}{4} \, C_d \frac{\rho_{\text{air}}}{\rho_{\text{quartz}}} d^{-1} \, (
\vec \veff - \vec u)|\vec \veff - \vec u|
    - \frac{g}{2 \alpha} \frac{\vec u}{|\vec u|}
    - g \, \vec \nabla \, h
    = 0,
\end{equation}
\noindent where $\vec \veff = v_{\text{eff}}\,\frac{\vec u_*}{|\vec u_*|}$.
$C_d$ is the drag coefficient of a grain.  $d_\ind{grain}$ and $\rho_\ind{grain}$ are the
diameter and the density of the grains.  $\alpha$ is a model parameter.

The sand flux due to saltation is then obtained by numerically solving the transport
equation:
\begin{equation}
  \label{eq:q}
  \text{div} \, \vec q = \frac{1}{l_s} q \left( 1 - \frac{q}{q_s} \right) \;
  \begin{cases}
    \Theta(h) & q < q_s\\
    1         & q \ge q_s
  \end{cases}
  ,
\end{equation}
with
\begin{equation}
  \label{eq:q_s}
  q_s = \frac{2 \alpha}{g}\,|\vec u| \left( |\vec \tau| - \tau_t \right) \quad \quad
  l_s = \frac{2 \alpha | \vec u|^2}{\gamma g} \, \frac{\tau_t}{(|\vec \tau| - \tau_t)}.
\end{equation}
Here $g$ is the gravity acceleration, and $\alpha$, $\beta$ and $\gamma$ are
model parameters taken from \cite{SauermannPhD}.

\paragraph{The time evolution of the surface:}

When the sand flux has been calculated, the height profile
is updated according to the mass conservation:
\begin{equation}
  \label{eq:surfupdate}
  \frac{\partial h}{\partial t} = - \frac{1}{\rhosand} \text{div} \, \vec q.
\end{equation}
\paragraph{Avalanches:}

In the last step, avalanches are simulated where necessary.  If the slope of
the sand surface exceeds the static angle of repose, sand is redistributed
according to the sand flux:
\begin{equation}
\label{eq:aval}
\vec q_{\text{aval}} = E\;(\tanh |\nabla h| - \tanh (\tan\theta_\text{dyn})\,\frac{\nabla h}{|\nabla h|}
\end{equation}
The surface is repeatedly changed according to (\ref{eq:surfupdate}) using this
flux, until the maximum slope is below the dynamic angle of repose,
$\theta_\text{dyn}$.  The hyperbolic tangent function serve only to improve convergence.

All these steps are repeated iteratively to simulate the evolution of the
shape.

\section{Results}

\begin{figure*}
\includegraphics[width=1.0\textwidth]{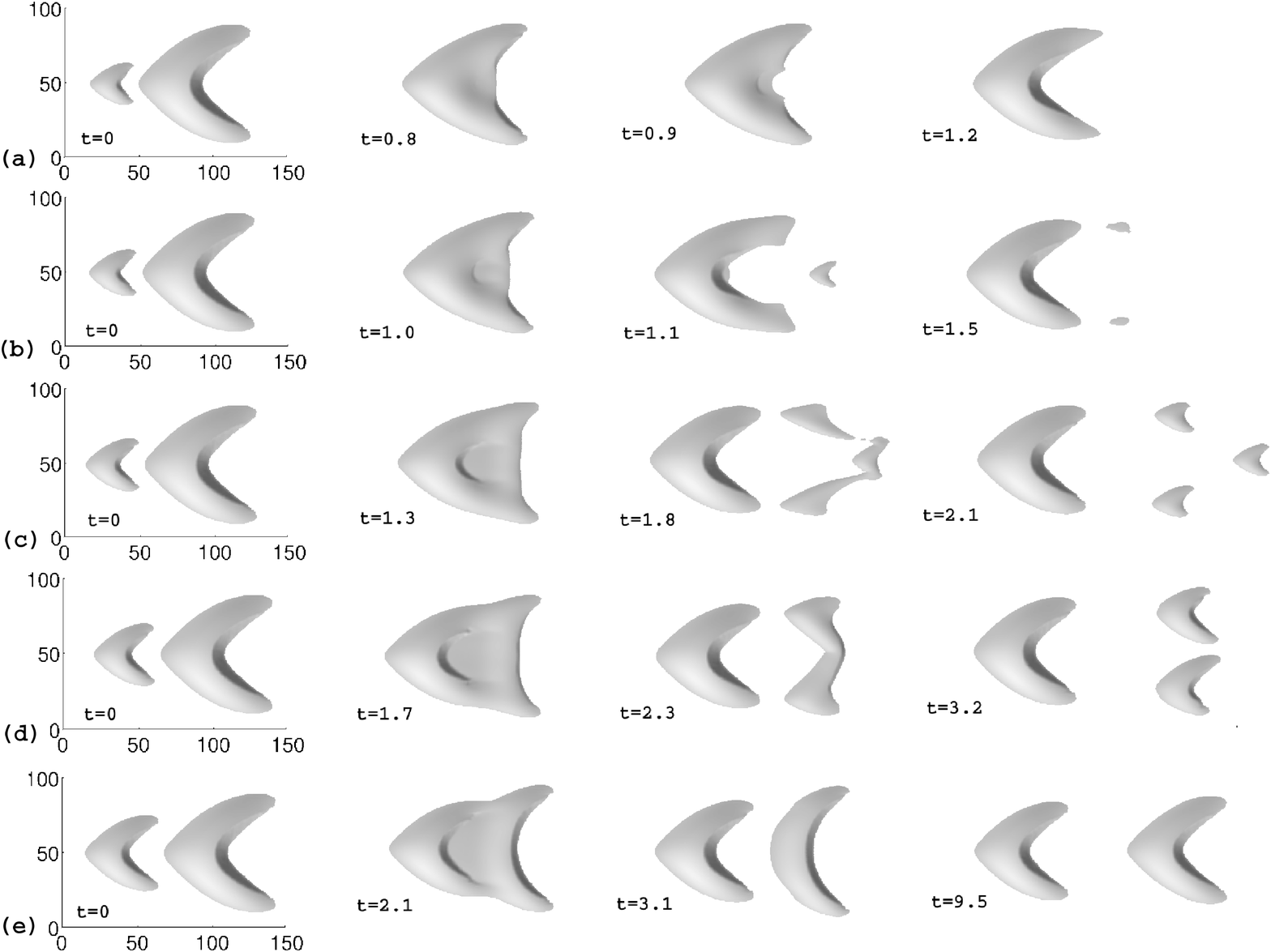}
\caption{Different situations during the collision of two Barchans for
$(V_h/V_H)_i=0.06$ (a), $0.08$ (b), $0.12$ (c), $0.17$ (d) and $0.3$ (e) using
open boundary condition. Coalescence (a), breeding (b) and (c), budding (d)
and solitary wave behaviour (e) take place. The time (t)
is in month. The initial volume and height of the big Barchan is
$6\times 10^3 m^3$ and $5\,m$ height, whereas the
heights of the smaller Barchan are $1.8\,m$ (a), $1.9\,m$ (b), $2.2\,m$ (c), $2.6\,m$ (d) and
$3.1\,m$ (e).}
\label{fig.2}
\end{figure*}

\begin{figure}[ht]
\includegraphics[width=0.94 \columnwidth]{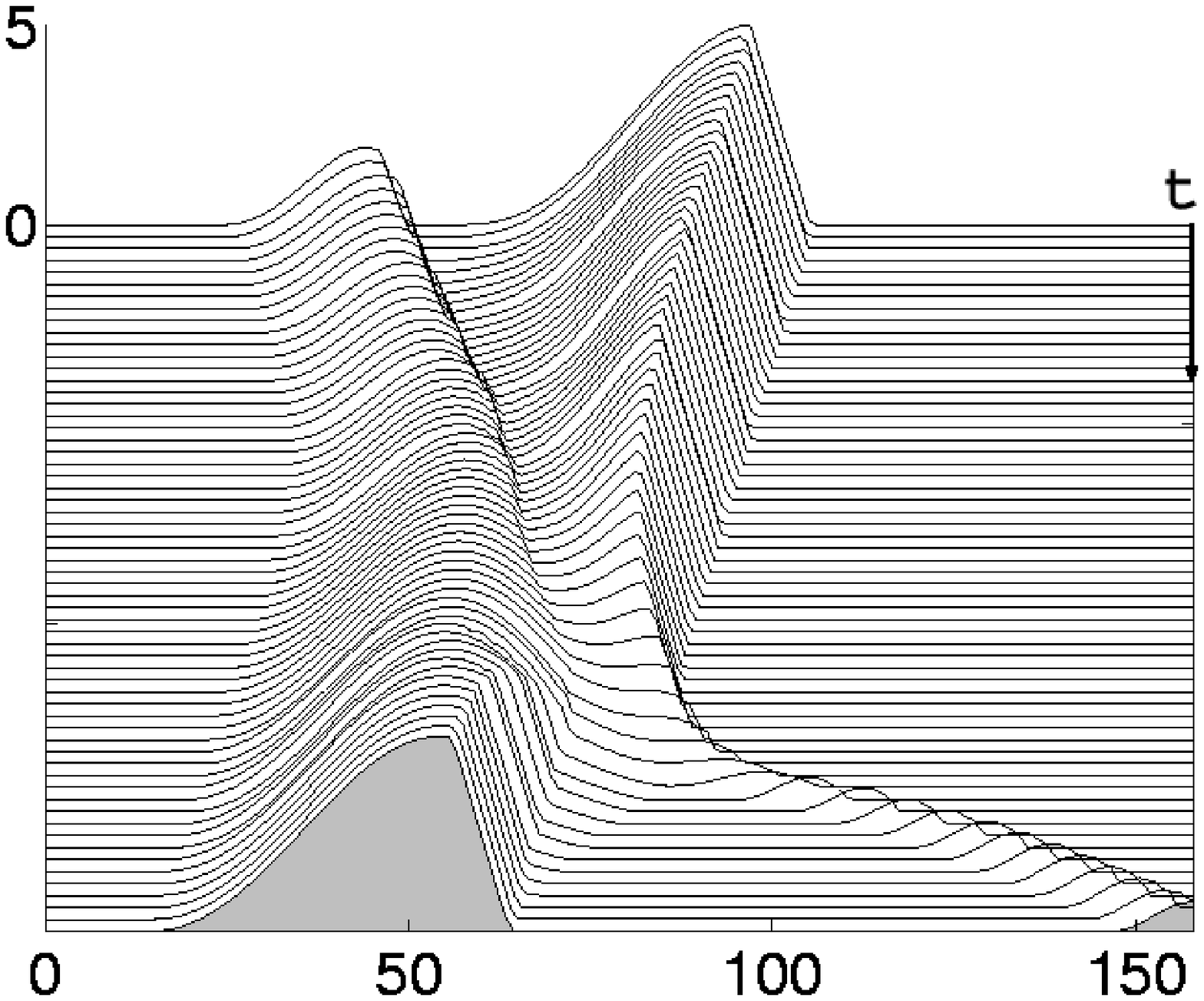}
\caption{Time evolution of the central slice of the `breeding' collision represented in the
figure ~\ref{fig.2}b. Note initially the small dune climbing on the biger one,
and finally their mass exchange that leads to their separation.}
\label{fig.3}
\end{figure}

We performed calculations by numerically solving the set of equations initially
placing a big Barchan (volume $V_H$) downwind of a smaller one
(volume $V_h$).  The strength of the wind blowing into the system is fixed to a shear
velocity of 0.5 m/s. The influx is 0.001 kg/ms, equal to the big Barchan equilibrium outflux. In
order to take into account the lack of scale invariance of Barchans, we
repeat the simulations for two different sizes of the big Barchan, with
initial volumes $V_H$: $6$ and $70$ $\times 10^3 m^3$.
The same general picture was observed. The smaller
Barchan at some point bumps into the larger one. This leads to a hybrid state
where the two dunes melt into a complex pattern. Four different situations can
be observed: coalescence (Fig.~\ref{fig.2}a),
breeding (Fig.~\ref{fig.2}b and ~\ref{fig.2}c),
budding (Fig.~\ref{fig.2}d) and solitary wave behavior (Fig.~\ref{fig.2}e)
depending only on the relative sizes of the two dunes. Thus, we chose as
control parameter the relative volume between the two dunes $V_h/V_H$.

The evolution of the hybrid state can be understood as the result of a
competition between two processes. The first one is the overlapping of both
dunes at the beginning of the collision that eventually can lead to
coalescence (Fig.~\ref{fig.3}, upper part). The second one is the effective mass exchange between the
dunes due to the changes induced to the wind shear stress due to the
approaching of both dunes.  In the hybrid state the wind shear stress over the
windward side of the bigger dune is reduced and thus, crest erosion is
enhanced. Besides, the wind shear stress over the lee side of the dune smaller
is also reduced but enhancing crest deposition. Thus, the dune smaller may
gain enough sand to become bigger than the one in front and therefore also
becomes slower. In this way, the before bigger dune can become the smaller one
and its velocity sufficiently large to leave the hybrid state. In this case the
dunes separate again (Fig.~\ref{fig.3}, bottom).

The collision process is crucially affected by the separation bubble, i.e.
the region after the brink and between the horns at which flow separation
occurs (see item $a.$ in the Model section). After the separation at the brink,
the flow streamlines reattach smoothly near the line segment whose end points
are the horns. There, sand transport continues again. However, inside the
separation bubble the flow is strongly reduced and, for simplicity, we set the
flux to zero. Hence, the upwind dune will absorb that part of the downwind dune
inside its separation bubble (Fig.~\ref{fig.2}c, ~\ref{fig.2}d and~\ref{fig.2}e).

For small relative volume ($V_h/V_H < 0.07$) both dunes coalesce to a single
one. In this case the relative velocity is high and hence the overlapping is
faster than their mass exchange. Small dunes have a short slip face which
disappears while climbing up the bigger one. This reduces the mass exchange and
leads to a complete absorption of very small Barchans (Fig.~\ref{fig.2}a). For larger
$V_h/V_H$ the slip face survives for longer time, mass exchange becomes relevant, and a small
barchan is ejected from the central part of the dune (Fig.~\ref{fig.3}). The
perturbation of the big dune shape, due to the overlapping of the small dune
behind, also propagates over the horns since there is no slip face. At the end of
each horn a
small dune is ejected. This phenomenon of triple ejection we call `breeding'. Fig.~\ref{fig.2}b
shows the snapshots. Note the qualitative similarity with the Barchan field
shown in Figure~\ref{fig.1}.

As the relative volume increases ($V_h/V_H > 0.14$), a smaller relative
velocity favours the mass exchange and reduces the overlapping process, leading
to the complete separation of both dunes. Nevertheless, the leaving dune lacks
 the central part of its windward side and cannot reach the stable Barchan
shape. Therefore, it splits into two new dunes, a phenomenon we call `budding'
(Fig.~\ref{fig.2}d). A similar phenomena was reported in experiments with 
sub-aqueous Barchans \cite{endo}.

Between the `breeding' and `budding', a transition occurs at which the `breeding' three ejected dunes
are connected forming a single dune that,
afterwards, splits into three again (Fig.~\ref{fig.2}c). For a higher relative volume the central
ejected dune decrease its size until disappear at `budding'.
We consider this transition also as `breeding'.

When $V_h/V_H$ is greater than $0.25$, the instability of the dune in front
disappears and both dunes develop to Barchans. Then, we observe solitary wave
behavior as shown in Figure~\ref{fig.2}e. In that case the dunes move with
similar velocity and the mass exchange is the main process of the evolution of
the hybrid state. The overlapping of both dunes is very small now and the
emerging dune looses merely a small fraction of its tail. Effectively it looks
as if the smaller dune just crosses the bigger one while in reality due to mass
exchange the two heaps barely touches each other.

The morphological phase diagram crucially influences the evolution of the number
of dunes in a Barchan field. In the coalescence region ($V_h/V_H < 0.07$) the
number of dunes decreases by one, whereas three dunes in the breeding 
($0.07 < V_h/V_H < 0.14$) and two dunes in the budding region ($0.14 < V_h/V_H < 0.25$)
appear, and the number of dunes increases by two and one respectively. 
Finally, in the solitary waves region ($0.24 < V_h/V_H$) a
small dune seems to cross the bigger one and the number of them remains
constant.

\begin{figure}[htb]
\includegraphics[width=0.94 \columnwidth]{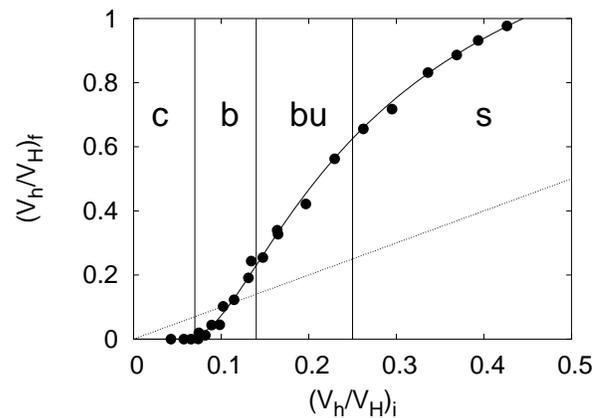}
\caption{Relation
between the relative volume of the dunes before ($(V_h/V_H)_i$) and after the
collision ($(V_h/V_H)_f$). The regions in the morphological phase
diagram are showed: coalescence (c), breeding (b), budding (bu) and solitary
waves (s). For coalescence only one dune emerges and thus, the relative volume
is zero, whereas for breeding, budding and solitary waves the relative volume seems to
follow a exp(-1/x) law (full line).}
\label{fig.4}
\end{figure}

In order to analyze the transition type between any two regimes, we introduce
two order parameters, the relative volume after the collision ($(V_h/V_H)_f$)
for the `coalescence-breeding' transition, and the inverse of time during which the
emerging dune does not split, for the `budding-solitary waves' transition. Note
that the former parameter is zero for coalescence, where no dunes emerge, and
the later one is zero for solitary wave behavior, where the leaving dune
evolves into a Barchan shape without splitting.

Figure~\ref{fig.3} shows the relative volume after the collision as a function
of the initial one, approximately related by the formula:

\begin{equation}
\left( \frac{V_h}{V_H}\right)_f \propto \exp{ \left(
\frac{-0.22}{\left(\frac{V_h}{V_H}\right)_i - r_c}\right)}
\end{equation}

\noindent   where $r_c = 0.03$. This suggests an essential singularity in the `coalescence-breeding'
transition. It implies
that  the volume of the leaving dune in the budding regime does not have a
minimum value, thus, after a Barchan it is  possible to find small dunes of
any height.

\begin{figure}[!htb]
\includegraphics[width=0.94 \columnwidth]{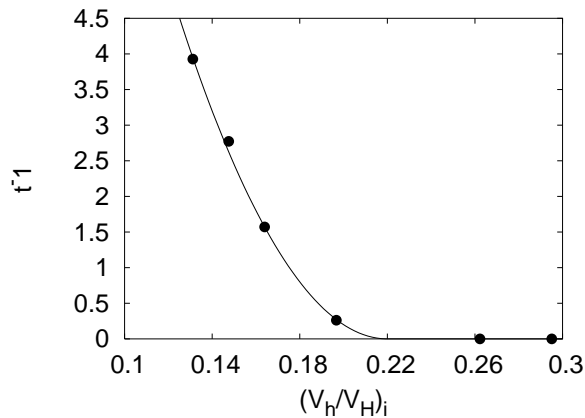}
\caption{Inverse of the time
during which the emerging dune does not split, as a function of the initial
relative volume. By definition, this time is infinitely long for the solitary
wave behavior. Note the continuity at the transition from budding to solitary waves,
suggesting a second order transition.}
\label{fig.5}
\end{figure}

On the other hand, Figure~\ref{fig.5} shows the inverse of the time during
which the emerging dune does not split. The continuity of the curve
at the `budding-solitary wave' transition suggest a second order transition.
Near the transition point the leaving dune splits in a infinite time. Thus, 
we cannot define in
a Barchan field a characteristic length for the budding process.

It is interesting to note that the relative volume increases after a solitary
wave collision  (Fig.~\ref{fig.4}). The collision process redistributes the
initial mass making both dunes more similar, giving rise to a size selection
mechanism in Barchan fields \cite{Hersen}.  However, due to the permanent
exchange of sand between the dunes, their sizes are not constant after the
collision and change with their influx and thus with the selected boundary
condition. Furthermore, the open problem of its stability \cite{Hersen}
introduces another uncertainty about their final sizes.

\section{Conclusions}

In this work we develop a morphological phase diagram showing that during the
collision of two Barchans four situations can be observed: coalescence,
breeding, budding and solitary waves (Fig.~\ref{fig.2}: a, b, c, d and e). At 
large scales the
collision process depicted here could lead to a selection of a characteristic
size of dunes in a Barchan field. However, we only considered perfectly aligned
dunes.

Calculations of very large dune fields are still difficult because of high
computational costs. One way out would be to consider a simplified model
containing the main features of dune movement and interaction. For that purpose
one could use the collision rules obtained in this work on a larger scale
\cite{Lima}. 

\acknowledgments

We thank E. Parteli and A. O. Sousa for stimulating discussions and Max Plank
Price and DAAD for financial support.

\end{document}